\documentclass[review]{elsarticle}

\usepackage{lineno,hyperref}
\usepackage{subfigure}
\modulolinenumbers[5]

\journal{Neurocomputing}

\begin{document}

\begin{frontmatter}

\title{Investigating Cultural Aspects in the Fundamental Diagram using Convolutional Neural Networks and Simulation}

\tnotetext[mytitlenote]{Thanks to Office of Naval Research Global (USA) and Brazilian agencies: CAPES, CNPQ and FAPERGS.}


\author[mymainaddress]{Rodolfo M. Favaretto\corref{mycorrespondingauthor}}
\cortext[mycorrespondingauthor]{Corresponding author}
\ead{rodolfo.favaretto@edu.pucrs.br}

\author[mymainaddress]{Roberto R. Santos}
\author[mymainaddress]{Marcio Ballotin}
\author[mymainaddress]{Paulo Knob}
\author[mymainaddress]{Soraia R. Musse}

\author[mysecondaryaddress]{Felipe Vilanova}
\author[mysecondaryaddress]{\^Angelo B. Costa}

\address[mymainaddress]{Virtual Human Simulation Laboratory VHLab -- Graduate Course on Computer Science, Pontifical Catholic University of Rio Grande do Sul, Porto Alegre, RS - Brazil}
\address[mysecondaryaddress]{Graduate Course on Psychology, Pontifical Catholic University of Rio Grande do Sul, Porto Alegre, RS - Brazil}

\begin{abstract}
This paper presents a study regarding group behavior in a controlled experiment focused on differences in an important attribute that vary across cultures - the personal spaces - in two Countries: Brazil and Germany. In order to coherently compare Germany and Brazil evolutions with same population applying same task, we performed the pedestrian Fundamental Diagram experiment in Brazil, as performed in Germany. We use CNNs to detect and track people in video sequences. With this data, we use Voronoi Diagrams to find out the neighbor relation among people and then compute the walking distances to find out the personal spaces. Based on personal spaces analyses, we found out that people behavior is more similar, in terms of their behaviours, in high dense populations and vary more in low and medium densities. So, we focused our study on cultural differences between the two Countries in low and medium densities. Results indicate that personal space analyses can be a relevant feature in order to understand cultural aspects in video sequences. In addition to the cultural differences, we also investigate the personality model in crowds, using OCEAN. We also proposed a way to simulate the FD experiment from other countries using the OCEAN psychological traits model as input. The simulated countries were consistent with the literature.
\end{abstract}


\end{frontmatter}


\section{Introduction}\label{sec:introduction}

Crowd analysis is a phenomenon of great interest in a large number of applications. Surveillance, entertainment and social sciences are fields that can benefit from the development of this area of study. Literature dealt with different applications of crowd analysis, for example counting people in crowds~\cite{Chan2009, cai2014}, group and crowd movement and formation~\cite{Zhou2014, jo2013review} and detection of social groups in crowds~\cite{Shao2014, Feng2015}. Normally, these approaches are based on personal tracking or optical flow algorithms, and handle as features: speed, directions and distances over time. Recently, some studies investigated cultural difference in videos from different countries using Fundamental Diagrams~\cite{Jelic:2012,Seyfried:FD:2005,Wu:2002,FLOTTEROD:2015,Shuchao:2017,CAO:2018}.

The Fundamental Diagrams -- FD, originally proposed to be used in traffic planning guidelines~\cite{Weidmann1993,Predtechenskii:1978}, are diagrams used to describe the relationship among three parameters: i) density of people (number of people per sqm), ii) speed (in meters/second) and iii) flow (time evolution)~\cite{Wu:2002}. In Zhang's work~\cite{Zhang:2012}, FD diagrams were adapted to describe the relationship between pedestrian flow and density, and are associated to various phenomena of self-organization in crowds, such as pedestrian lanes and jams, such that when the density of people becomes really high, the crowd stops moving. It is not the first time cultural aspects are connected with FD. Chattaraj and his collaborators~\cite{Chattaraj:2009} suggest that cultural and population differences can also change the speed, density, and flow of people in their behavior.

Favaretto and his colleagues discussed cultural dimensions according to Hofstede analysis~\cite{Hofstede:2011} and presented a methodology to map data from video sequences to the dimensions of Hofstede cultural dimensions theory~\cite{Favaretto:2016} and also a methodology to extract crowd-cultural aspects~\cite{favaretto:2017} based on the Big-five personality model (or OCEAN)~\cite{costa07}. In his work, Favaretto~\cite{favaretto:2017} proposed a way to map geometrical features (such as speed, angular variation and distances) from pedestrians tracking to OCEAN dimensions. 

In this paper, we want to investigate cultural aspects of people when analyzing the result of FD among two different Countries: Brazil and Germany. We used the Pedestrian Fundamental Diagram experiment performed in Germany and perform the experiment in Brazil, in order to compare these two different populations. Our goal is to investigate the cultural aspects regarding distances in personal space analyses. FD was chosen since the populations are performing the same task in a controlled environment with same amount of individuals. We also propose a way to simulate other countries using OCEAN as input to generate geometrical features (such as speed, angular variation, etc.) of each pedestrian. The next section discusses the related work, and in Section~\ref{sec:approach} we present details about the proposed approach with a statistical analysis (Section~\ref{sec:results}), followed by the discussion and final considerations in Section~\ref{sec:conclusion}.

\section{Related Work}
\label{sec:related}

Cultural influence can be considered in crowds attributes as personal spaces, speed, pedestrian avoidance side and group formations~\cite{Fridman2013}. Personal space refers to the preferred distance from others that an individual maintains within a given setting. This area surrounding a person’s body into which intruders may not come is the personal space~\cite{sommer:1969}. It serves mainly to two main functions: (i) communicating the formality of the relationship between the interactants; and (ii) protecting against possible psychologically and physically uncomfortable social encounters
~\cite{aiello:1980}. People from various cultural backgrounds differ with regard to their personal space~\cite{baxter:1970}. These differences reflect the cultural norms that shape the perception of space and guide the use of space within different societies~\cite{hall:1966}.

Recently, a study on personal space employing a projective technique was conducted in 42 countries \cite{sorokowska:2017}. Participants had to answer a graphic task marking which distance they would feel comfortable when interacting with: a) a stranger, b) an acquaintance, and c) a close person. This way the authors could evaluate the projected metric distance for a) social distance, b) personal distance and c) intimate distance. The number of countries assessed in the study of Sorokowska and colleagues \cite{sorokowska:2017} promote conclusions from different cultures and indicated some new possible categorization of the cultures but also to design objects or implement changes in the real world. 
The project of public transportations, for example, can be improved by the analysis of real personal space in different countries, since the invasion of the personal space in trains elicits psychophysiological responses of stress~\cite{evans_wener:2007}. Furthermore, the project of human-robots has also been improved through the analysis of personal space~\cite{walters:2009}. As it is important that robots do not invade the personal space of its users, the configuration of its distances might benefit from studies that employ analysis of daily preferred interpersonal distances across different countries.

Our idea here is to identify different aspects among populations from Brazil and Germany regarding distances in individual's personal space. However, differently from the projective technique proposed by~\cite{sorokowska:2017}, we want to use video sequences, real populations and computer vision techniques to proceed with personal space analyses. Next section presents the methodology adopted to detect and track the individuals in the experiment and how we perform the statistic information extraction.

\section{The proposed approach}
\label{sec:approach}

\begin{figure}[h]
\centering
\includegraphics[width=11cm]{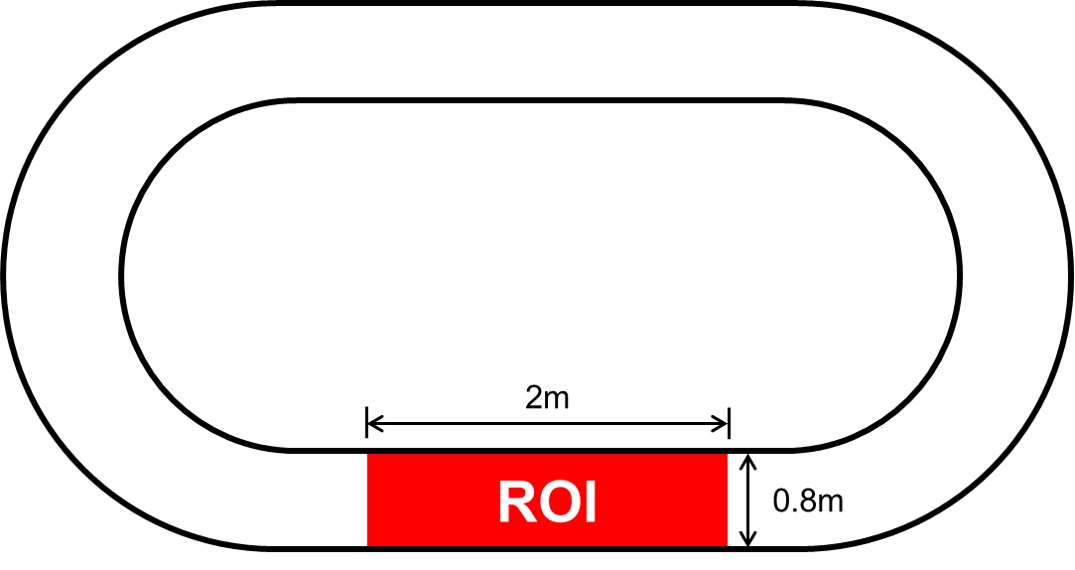}
\caption{Sketch of the FD experimental setup according to \cite{Chattaraj:2009}.}
\label{fig_config_exp}
\end{figure}

We propose a 2-step methodology responsible for trajectories detection and statistical data extraction/analysis. The first part aims to obtain the individual trajectories of observed pedestrians in real videos using machine learning algorithms. We performed the Fundamental diagram experiment in Brazil, as illustrated in Figure~\ref{fig_config_exp}.

This experiment in Brazil was conducted as described in~\cite{Chattaraj:2009}. With the same populations (N=1, 15, 20, 25, 30 and 34) and physical environment setup. In addition, we obtained from Germany~\footnote{We have access to such videos thanks to the authors of database of PED experiments, available at \url{http://ped.fz-juelich.de/db/}.} video with populations (N=1, 15, 25 and 34), so N=20 and 30 were not used in our analysis.

The corridor was built up with markers and tape on the ground. Its size and shape is presented in Figure~\ref{fig_config_exp}. The length of the corridor is $l_{corr} = 17.3m$. The width of the passageway is $w_{corr} = 0.8m$, which is sufficient for a single person walk.
In addition, we can observe a rectangle of 2 x 0.8 meters which illustrates the Region of Interest (ROI) where the populations were captured to be analyzed, as proposed in~\cite{Chattaraj:2009}.

\begin{figure}[!h]
\centering
\subfigure[fig:fdbra15][Brazil]{\includegraphics[width=10cm]{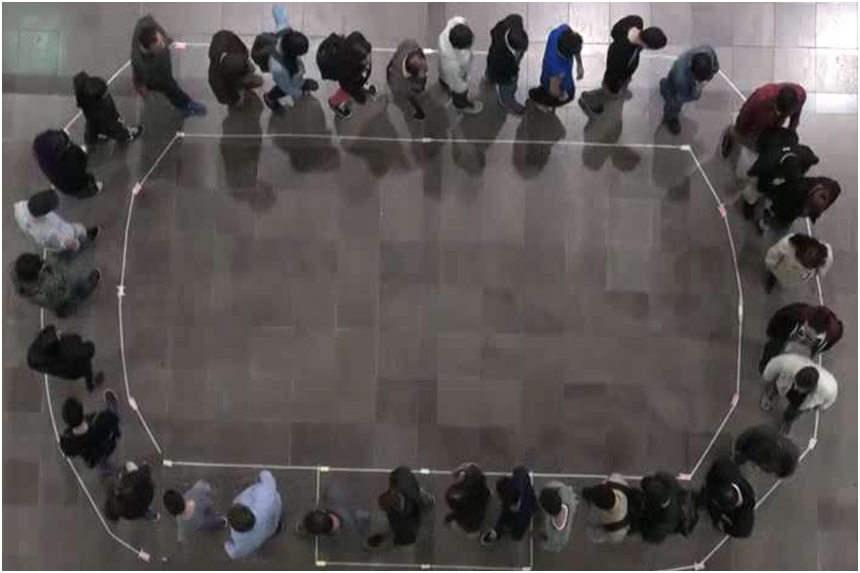}} \\
\subfigure[fig:fdger15][Germany]{\includegraphics[width=10cm]{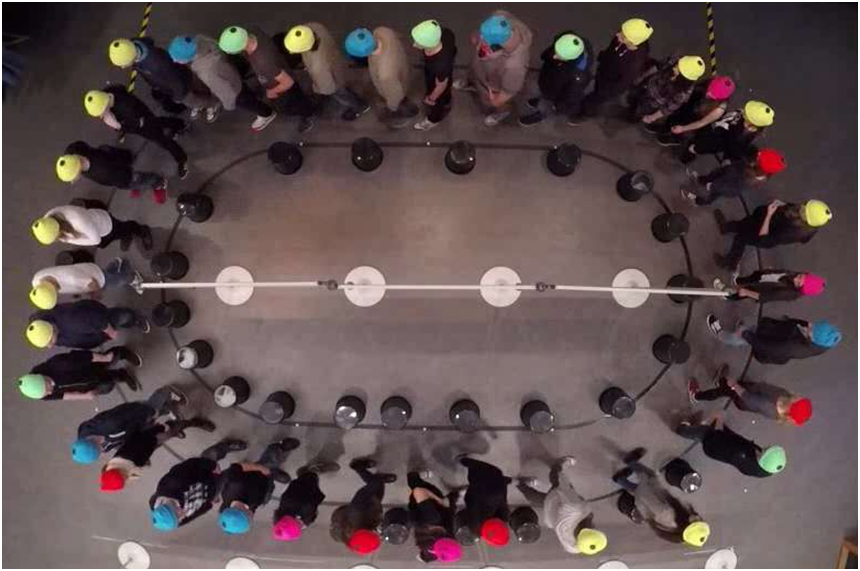}}
\caption{Some pictures extract from the experiment: (a) performed in Brazil with $N=34$ and (b) performed in Germany with $N=34$.}
\label{fig_FD_exp}
\end{figure}

For the experiment, 
the camera was positioned in the top, eliminating the video perspective. All the individuals were initially uniformly distributed in the corridor. After the starting instruction, every individual should walk around the corridor twice and then leave the environment while keep walking for a reasonable distance away, eliminating the tailback effect. Figure~\ref{fig_FD_exp} shows the experiment performed in Brazil and Germany, with $N=34$ (where $N$ is the number of people).

In the first step of our method, the people detection and tracking is performed using Convolutional Neural Networks (CNNs). In the second step, the statistical information is obtained from trajectories and analyzed in order to find neighbor individuals and compute distances among them. These modules are presented in sequence.

\subsection{People detection and tracking}
\label{subsec:tracking}

\begin{figure}[!h]
\centering
\subfigure[fig:fdbra15][Test using VOC]{\includegraphics[width=10cm]{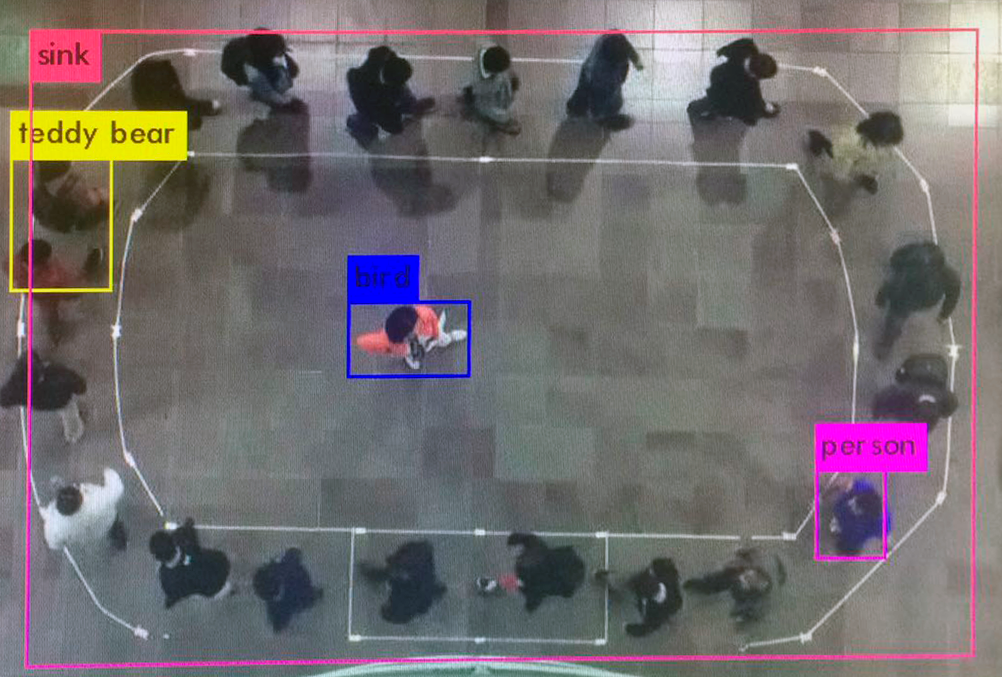}} \qquad
\subfigure[fig:fdger15][Training Results]{\includegraphics[width=10cm]{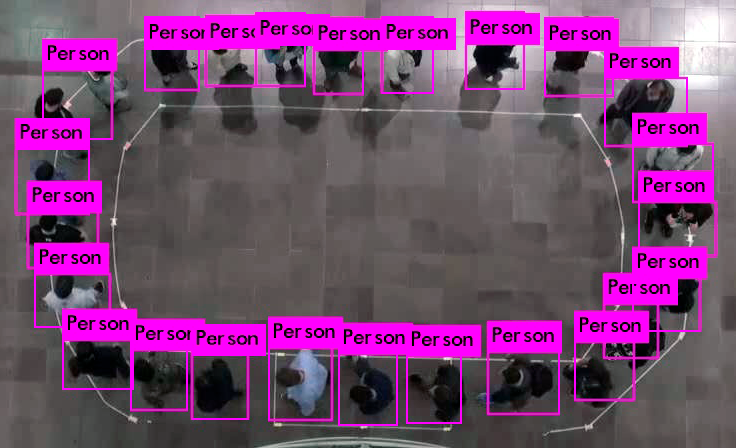}}
\caption{Test using VOC and trained pattern configuring (a). Training Results in a video from Brazil (b).}
\label{fig_detect}
\end{figure}

Since our goal was to accurately track the issues involved in the FD experiment, we decided to use the recent convolutional neural networks (CNNs). We use the real-time detection framework, Yolo with reference model Darknet~\cite{redmon2016yolo9000}. Initially, we used trained models with public datasets, named COCO ~\cite{DBLP:journals/corr/LinMBHPRDZ14} and PASCAL VOC~\cite{Everingham15}. However, due to very different camera position in the video sequences, the tracking did not work well, as can be seen in Figure~\ref{fig_detect}(a).

So, we proceed with a dataset generation to be used for the network training. We used the videos with 20 and 30 people performed in Brazil. We choose this two experiments (with 20 and 30 pedestrian) for training process because we do not have the corresponding amount of people from the Germany dataset. We included in the training dataset one image at each 50 frames, resulting in 45 images for movie with 20 people and 83 for video with 30 people. Table~\ref{tbl:qtdeTrain} shows the number of images used in training, validation and testing phases. Obtained accuracy in our method for videos from Brazil was 98.2 \% with 15 people, 98.4 \% with 25 people and 97.8 \% with 34 people. Table~\ref{tbl:result} demonstrates the accuracy of both Countries in the respective videos.

\vspace{1.5mm}
\begin{table}[ht]
\centering
\caption{Configuration of the Dataset used in the experiment.}
\label{tbl:qtdeTrain}
\begin{tabular}{c|c|c|c}
\hline\noalign{\smallskip}
Goal & Images & Annotations & Country \\
\noalign{\smallskip} \hline
\noalign{\smallskip} \hline \noalign{\smallskip} 
Train & 128 & 3833 & Brazil\\
Valid  & 96 & 1536 & Brazil\\
Test - 15 people & 1596 & 23530 & Brazil \\
Test - 25 people & 3124 & 73250 & Brazil \\
Test - 34 people & 5580 & 178448 & Brazil \\
Test - 15 people & 2372 & 71846 & Germany \\
Test - 25 people & 3322 & 74005 & Germany \\
Test - 34 people & 3504 & 110500 & Germany \\
\noalign{\smallskip} \hline
\end{tabular}
\end{table}

\vspace{1.5mm}
\begin{table}[h]
\centering
\caption{Accuracy (\%) obtained.}
\label{tbl:result}
\begin{tabular}{c|c|c|c}
\hline\noalign{\smallskip}
Country & 15 people & 25 people & 34 people \\
\noalign{\smallskip} \hline
\noalign{\smallskip} \hline \noalign{\smallskip} 
Brazil & 98.2\% & 98.4\% & 97.8\% \\
Germany & 93.0\% & 92.3\% & 91.0\% \\
\noalign{\smallskip} \hline
\end{tabular}
\end{table}

\subsection{Statistical Data Extraction and Analysis}
\label{sec:statistic}

As a result of tracking process, described in last section, we obtained the 2D position $\vec{X_i}$ of person $i$ (meters), at each timestep in the video. Positions are used to compute the Fundamental Diagram.
We adopted the already used hypothesis~\cite{Jacques2007} to approximate the personal space using a Voronoi Diagram (VD). Indeed, we use the output of VD to compute the neighbor of each individual in order to calculate the pairwise distances. As our pedestrian tracking could not be applied to find out the order of pedestrians in the video (we do not know the order in which the pedestrians were tracked, e.g. $i$ and $i+1$), we use the output of the Voronoi Diagram to compute the neighbour of each individual (pedestrians in front and behind) to calculate distances between each pedestrian and his/her predecessor. So, the distance between individual $i$ and the one in front of him/her $i+1$ is considered the personal space of $i$, in this work. So, we compute such distances in the ROI, at the first moment the second individual entries in the ROI illustrated in Figure~\ref{fig_config_exp}.

\begin{figure}[h]
\centering
\includegraphics[width=12cm]{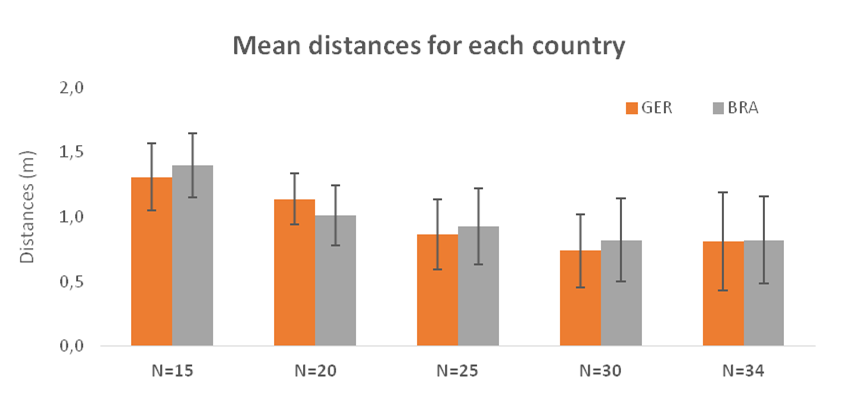}
\caption{Mean personal distances observed in each population.}
\label{fig_mean_dis}
\end{figure}

\begin{figure}[h]
\centering
\includegraphics[width=12cm]{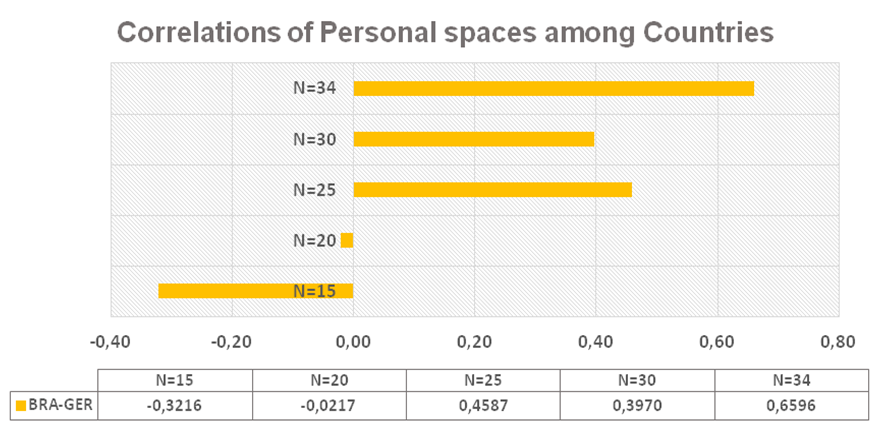}
\caption{Correlations of personal space among the countries.}
\label{fig_corr_dis}
\end{figure}

Once we have computed all personal spaces for all individuals from the two populations, we conducted the following analysis. First, we show in Figure~\ref{fig_mean_dis} the mean distances observed in each population. As expected, the personal space reduces as the density increases.
The correlations of distances among the two populations are shown in Figure~\ref{fig_corr_dis}.

\vspace{1mm}

As can be observed in Figure~\ref{fig_corr_dis}, the Pearson's correlations among the populations increase as the densities increase too. Based on this affirmation, our hypothesis is that in high densities, people act more as a mass and less as individuals~\cite{vilanova2017}, which ultimately affects behaviors according to their own culture. This assumption is coherent with one of the main literatures on mass behavior~\cite{LE_BON_THE_CROWD}.

\begin{figure}[h]
\centering
\subfigure[fig:fdbra15][$N=15$]{\includegraphics[width=2.35in]{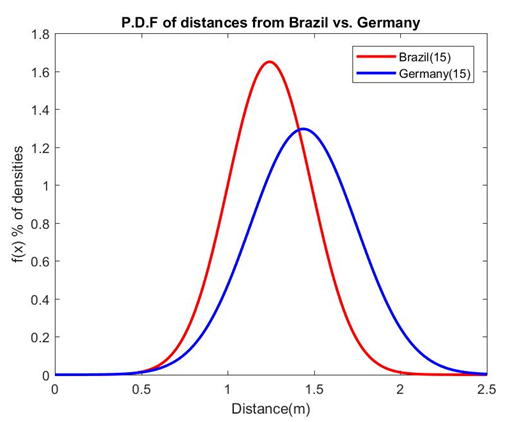}}
\subfigure[fig:fdger15][$N=25$]{\includegraphics[width=2.35in]{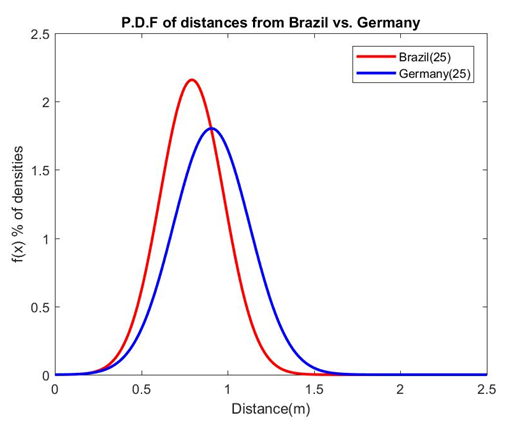}}
\subfigure[fig:fdger15][$N=34$]{\includegraphics[width=2.35in]{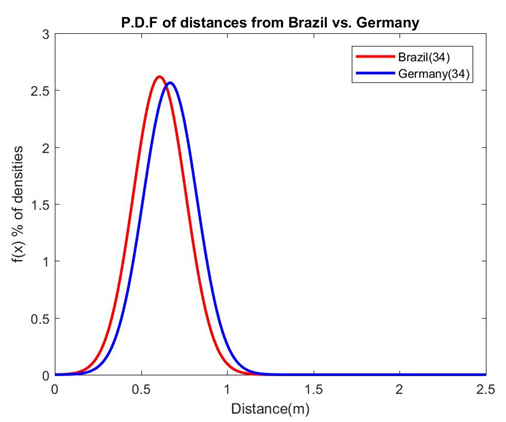}}
\caption{Probability distribution function (PDF) from the distances between the individuals in the experiment with, respectively: (a) $N=15$, (b) $N=25$ and (c) $N=34$.}
\label{fig_pdf}
\end{figure}

Figure~\ref{fig_pdf} shows an analysis of the Probability Distribution Function (PDF) applied on the personal spaces. The three plots represent the probability of distributions for each observed personal space in the interval $[0-2.5]$ meters. The red lines represent the probabilities from Brazil while the blue line represents the probabilities from Germany. The individuals from Germany keep a higher distance from each other than individuals from Brazil.

\begin{figure}[h]
\centering
\includegraphics[width=3.5in]{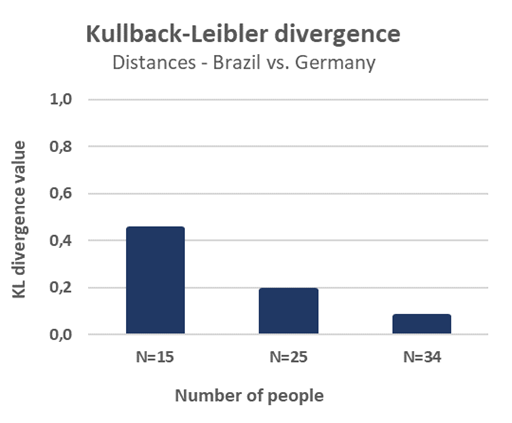}
\caption{Kullback-Leibler divergence from the distributions of distances in Figure \ref{fig_pdf}.}
\label{fig_kld}
\end{figure}

The distances performed by Brazilian individuals seems to have a lower standard deviation than distances performed by individuals from Germany (the width of the Gaussian curve is smaller in Brazil). The distances from the individuals in both countries gets more similar (the red and the blue lines are more similar when $N=34$ than $N=15$), corroborating with the mass idea. Also in Figure~\ref{fig_kld}, we present the Kullback-Leibler divergence from the probability distribution of distances among the countries. The Kullback–Leibler (KL) divergence~\cite{Kullback} (also called relative entropy) is a measure of how one probability distribution diverges from a second. It is interesting to see that as the density increases, the KL divergence decreases.

\subsection{Simulating Fundamental Diagram}
\label{sec:simulation}

In this section we describe our proposal to simulate the Fundamental Diagram. Our idea is to simulate FD experiments with varied populations. Once in last section we analyzed the FD in two Countries, our main goal here is to investigate if we can simulate FD for other Countries in a coherent way, if compared with the literature. That is why we chosen OCEAN (Openness, Conscientiousness, Extraversion, Agreeableness and Neuroticism) psychological traits model, proposed by Goldberg~\cite{OCEANEmergentTrait} to serve as input in our method. In addition, it is has been already used in the context of simulation. For instance, Durupinar et al.~\cite{durupinar2016psychological} developed a simulation model based on psychological traits aiming to represent emotions and emotion contagion between agents in an effective way.

Therefore, there is a specific literature presenting the OCEAN of different Countries~\cite{OCEAN_GLOBAL_COMPARISON} that can inform input values in our method. As mentioned before, Favaretto et al.~\cite{favaretto:2017}  comprehends equations to map pedestrian behavior, from video sequence, to OCEAN individual values. So, we extended this model to propose a way to, having the OCEAN as input, find out geometrical information regarding how people evolve in simulations. We decided to use three parameters to simulate FD, that are achieved based on equations proposed by Favaretto~\cite{favaretto:2017}: \textit{collectivity}, \textit{angular variation} and \textit{linear speed}. 

Collectivity is related to the group cohesion, i.e. the higher is the cohesion more collective behaviors the population has~\cite{lang1987}. According to Dyaram et al.~\cite{dyaram2005unearthed}, members of a strongly cohesive group tends to stay together, not leaving the group, as well to be an active part of it. 
Angular variation is computed as a function of the vector that represents the goal direction of agent $i$ and the third parameter represents the linear speed of $i$. These three parameters were proposed in Favaretto~\cite{favaretto:2017} and are inversely mapped, in this work, to be computed having OCEAN values as input, as described in following equations. Equation~\ref{eq:method:feature:collectivity} describes collectivity $\phi_i$ of agent $i$ as a function of $E_i$, $A_i$ and $N_i$ that state for some of input OCEAN parameters for $i$:

\begin{eqnarray}
\phi_i &=& \frac{2\frac{A_i}{100} + \frac{50}{8N_{i} - 100} + 2\frac{E_{i}}{100} + 2( 1 - \frac{N_{i}}{100})}{7}. \label{eq:method:feature:collectivity}
\end{eqnarray}

Equation~\ref{eq:method:feature:ang_var} describes the angular variation of $i$ as a function of $O_i$, $A_i$ and $\phi_i$ parameters:

\begin{eqnarray}
\alpha_i &=& \frac{ 1 - \frac{O_{i}}{100} +  1.208 - \frac{1}{16\phi_i} - \frac{E_1}{100}}{2}.
\label{eq:method:feature:ang_var}
\end{eqnarray}

And Equation~\ref{eq:method:feature:speed} refers to linear speed of agent $i$ and it is impacted by $E_i$, $C_i$ and $\alpha_i$ parameters:

\begin{eqnarray}
s_i &=& \frac{\frac{0.04C_{i} - (4\alpha_{i})^{-1}}{4} + \frac{\frac{E}{100} - \alpha_i + 1}{2}} {2} .
\label{eq:method:feature:speed}
\end{eqnarray}

Table~\ref{tb:method:feature:sum} shows a summarization of the relations among OCEAN and geometric parameters. Collectivity is related to Extraversion, Agreeableness and Neuroticism traits; Angular Variation is related to agent Openness, Extraversion and cohesion, and finally Speed is dependent on Consciousness, Extraversion and angular variation.

\begin{table}[ht]
\renewcommand{\arraystretch}{1.45}
\centering
\caption{Relationship between the agent features and input OCEAN dimensions.}
\label{tb:method:feature:sum}
\begin{tabular}{c|c}
    \hline\noalign{\smallskip}
  Agent Features  &  Related input \\
	\noalign{\smallskip} \hline \noalign{\smallskip}
    \hline\noalign{\smallskip}
  Collectivity ($\phi_i$)  & E, A, N \\ 
  Angular Variation ($\alpha_i$)  & O, E, $\phi_i$ \\ 
  Speed ($s_i$) & C, E, $\alpha_i$ \\
  \noalign{\smallskip} \hline
\end{tabular}
\end{table}

It is important to notice that the Extraversion trait is related to all features.
As mentioned in Favaretto~\cite{favaretto:2017}, the Extraversion trait comprehends the majority of items related to crowd behavior, so, being necessary for all equations as proposed in the present work. 

We use $\phi_i$, $\alpha_i$ and $s_i$ of agent $i$ to impact its motion in FD. Virtual humans are modeled to move in a pre-defined order in FD scenario having $\alpha$ and $s$ as angular and linear speed, respectively. The agent collectivity is used to define the cohesion of the group which the agent is a member.
A group's cohesion will be calculated as the mean value of its participants collectivity factor $\phi^B$.
Groups with cohesion close to one, have stronger bounds between their participants and will be harder to separate over the simulation conditions and the opposite is true for cohesion value close to zero.

In our method, a cohesion value~$\zeta_g$ is set to define how much a group~$g$ tends to stay together, in the interval~$[0,3]$, where~$0$ is the lowest cohesion value and~$3$ is the highest. This interval was defined according to the work proposed by Favaretto et al.~\cite{Favaretto:2016}, in which authors extract groups parameters from video sequences. Furthermore, a cohesion distance value~$\mu_g$ is defined to represent the maximum distance an agent can be away from the rest of the group~$g$, without leaving it and break the group structure. This cohesion distance is calculated as follows in Equation~\ref{eq:cohesion:distance}:

\begin{equation}
\label{eq:cohesion:distance}
\mu_g = H_s - (\zeta_g (\frac{H_s - H_p}{\zeta_{max}})),
\end{equation} 

\noindent where~$H_p$ is the Hall's personal space and~$H_s$ is the Hall's social space. This distance spaces are described by Hall~\cite{PROXEMICS_GLOBAL_COMPARISON} which defines regions that a person feels comfortable to maintain at each specific personal or social levels.
$\zeta_{max}$ value stands for Maximum Cohesion ($\zeta_{max}=3$) and represents the higher cohesion value a group can achieve.

For instance, if~$\zeta_g=0$ for a certain low cohesive group~$g$, then~$\mu_g=3.6$ meters, i.e. this group can have its members more spread in the environment.
On the other hand, if~$\zeta_g=3$ then~$\mu_g=1.2$ meters, meaning that members, in a more cohesive group, stay close to each other in order to be a group, since they have a strong connection and are more attracted to each other. 

\section{Experimental Results}
\label{sec:results}

In this section we present results about FD investigation firstly based on video sequences, then based on simulations.

\subsection{FD in video sequences compared with the Literature}

We performed a comparison among the preferred distance people keep from others evaluated in a study performed by Sorokowska~\cite{sorokowska:2017} and the results obtained from the experiment performed in our approach. In the Sorokowska's work, the answers were given on a distance (0-220 cm) scale anchored by two human-like figures, labeled A and B. Participants were asked to imagine that he or she is Person A. The, the participant was asked to rate how close a Person B could approach, so that he or she would feel comfortable in a conversation with Person B.

\begin{figure}[h]
\centering
\subfigure[soro15][When $N=15$]{\includegraphics[width=6cm]{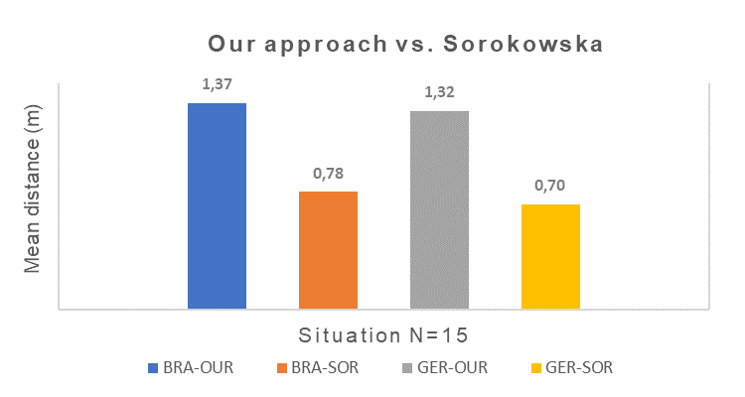}}
\subfigure[soro20][When $N=20$]{\includegraphics[width=6cm]{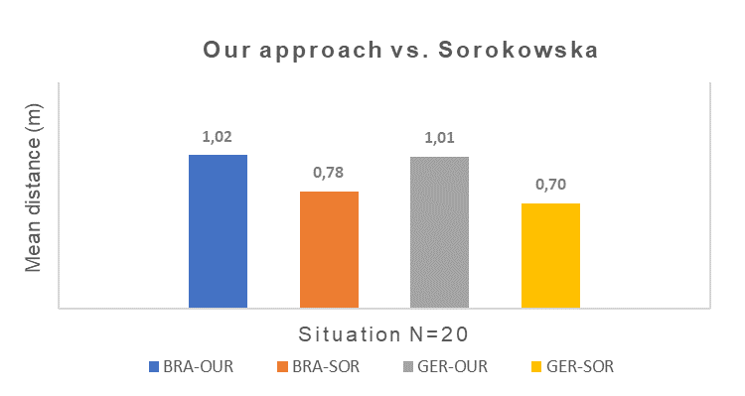}} \\
\subfigure[soro25][When $N=25$]{\includegraphics[width=6cm]{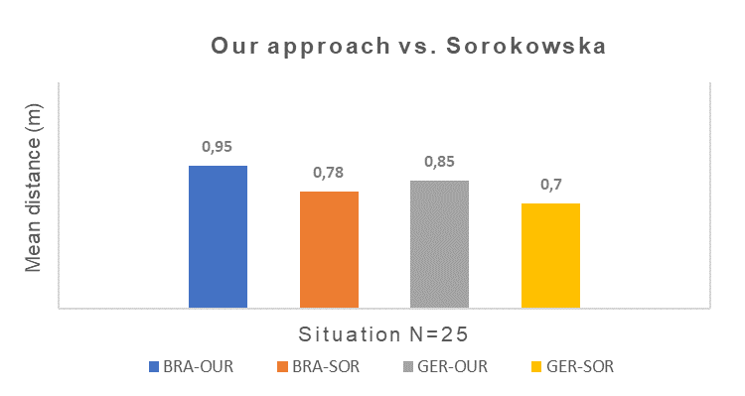}}
\subfigure[soro30][When $N=30$]{\includegraphics[width=6cm]{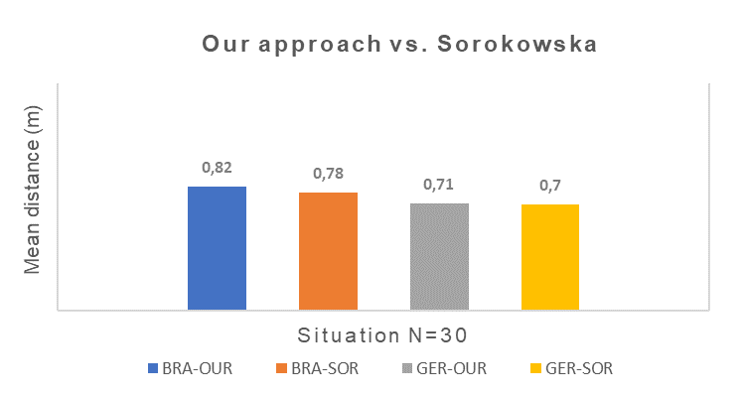}}
\caption{Our approach versus Sorokowska \cite{sorokowska:2017} with different number of people in the experiment.}
\label{fig_our_vs_soro}
\end{figure}

Figures~\ref{fig_our_vs_soro} show the comparison of four different FD scenarios, containing respectively 15, 20, 25 and 30 people. In our approach we measure the distances a person A keeps from a person B right in front of he or she. As said before, we used VD to determine which person is the neighbor of the other. For the comparison, in the Sorokowska's approach we select the evaluation from acquaintance people, where the people are not close neither strangers, similar to people in our experiment.  

As we can see in Figure~\ref{fig_our_vs_soro}, in spite of the fact that distances from our approach are higher than the ones from Sorokowska, the proportion is similar in all the scenarios. People from Brazil keeps higher distances from others than people from Germany (according to our approach, in the $N=15$ configuration, people from Brazil are about 0.5m more distant from each other than in Germany, while in the Sorokowska approach, people from Brazil are 0.8m more distant). It's interesting to notice that as the number of people increases, more similar to the values obtained by Sorokowska it gets (When $N=30$ the values are quite similar). Although they are different experiments, our method proves in a real scenario that people actually behave according to the preferences answered in Sorokowska's research.

\subsection{FD in simulations compared with the Literature}

We modeled the fundamental diagram experiment similar to the measures of Chatarraj et al.~\cite{Chattaraj2009}, using the BioCrowds simulator~\cite{Bicho2009}. In BioCrowds, each agent in the environment perceives a set of markers (dots) on the ground (described through space subdivision) within its observational radius and moves forward to its goal based on such markers (unoccupied and closest to the agent than any other one). This is the main feature of the BioCrowds simulator, which supports main behaviours observed from crowd simulations (e.g., lanes and arcs formation). 

\begin{figure}[ht]
  \centering
  \includegraphics[width=8cm]{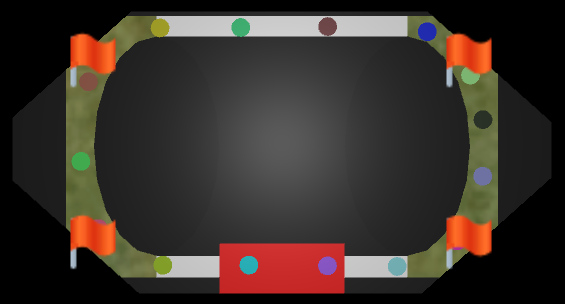}
  \caption{Example of Fundamental diagram experiment in BioCrowds with goals and $15$ agents.}
    \label{fig:fd:setup:unity}
\end{figure}

As output, BioCrowds measures the position of each agent at each frame, similar the tracking process performed with the FD videos. For more information on BioCrowds, please refer to~\cite{Bicho2009}. In this work we simulated an FD using BioCrowds for the same population tested using the similar environmental setup, as described in~\cite{Chattaraj:2009},  adding goal (represented as red flags) at every corner of the scenario, as shown in Figure~\ref{fig:fd:setup:unity}.

The agents are programed to seek the next goal anticlockwise, this way they keep looping. Knowing the agent in front of it, we are able to calculate the Euclidean distance between them, this distance is the personal distance of this agent. With this method we are able to determinate the personal distance of every agent during the simulation.

\begin{figure}[ht]
  \centering
  \includegraphics[width=8cm]{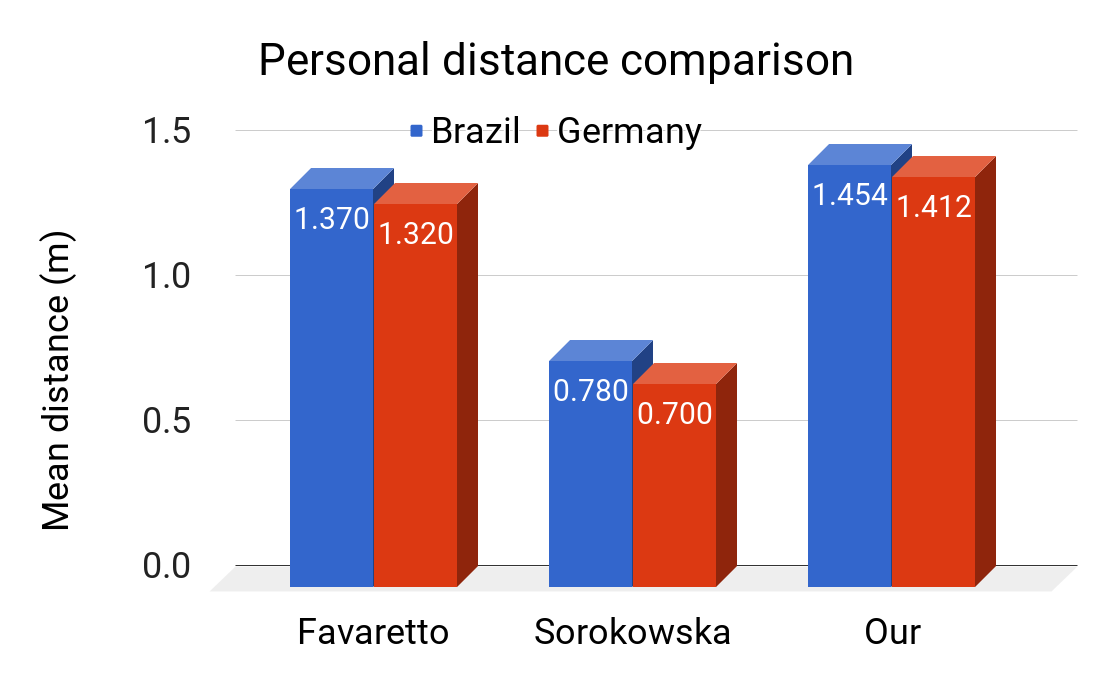}
  \caption{Comparison between the personal distances found by Favaretto et al. videos~\cite{favaretto:2017}, Sorokowska study~\cite{sorokowska:2017} and our method for Brazilian and German cultures.}
    \label{fig:fd:result}
\end{figure}

Based on Favaretto et al.~\cite{favaretto:2017} experiment, we executed simulations using the OCEANs presented by McCrae~\cite{OCEAN_GLOBAL_COMPARISON} for Germans and Hispanic Americans groups. For Germany, the used inputs are:~$O=56.7$,~$C=46.7$,~$E=47.3$,~$A=49.1$ and~$N=52.8$. For Brazil, we assumed the Hispanic Americans values:~$O=51.2$,~$C=51.6$,~$E=47.5$,~$A=47.1$ and~$N=49.5$. With this setup, we collected the personal distances of all agents during the simulation and calculated the mean personal distance value of all agents. Figure~\ref{fig:fd:result} shows a comparison chart between the results obtained by each study for both cultures.

Though obtained values are different for every approach, we observe a similar behavior in all of them, Brazilian personal distances are slightly greater than Germans in all approaches, i.e. in the video analysis (as presented by Favaretto et al.~\cite{favaretto:2017}), in social literature (as described by Sorokowska et al.~\cite{sorokowska:2017} ) and indicates that our method can be comparable to the real life.

This way, our method successfully represented the proxemics of both cultures. About the difference of the obtained values, we believe the small difference obtained between Favaretto's et al.~\cite{favaretto:2017} and our work results can be explained due the involved simulation parameters. These parameters would require extra tuning to represent the reality with more accuracy as we used literature to set the majority of parameters, e.g. the agent markers-detection radius~$R_i$ and the human average walk speed~$s_{avg}$.

Also in the experiment conducted by Sorokowska~\cite{sorokowska:2017}, the individuals were asked to answer their comfort distances in an image, in a survey and maybe the difference indicates that Physical space is not accurate with virtual abstractions. Although the results show a similar behavior, the interviewed individuals laked in visual and sensorial informations that could make them feel uncomfortable in a way they feel not by answering it on a paper, but we reinforce the behaviors are indeed similar.

\begin{figure}[t]
  \centering
  \includegraphics[width=\linewidth]{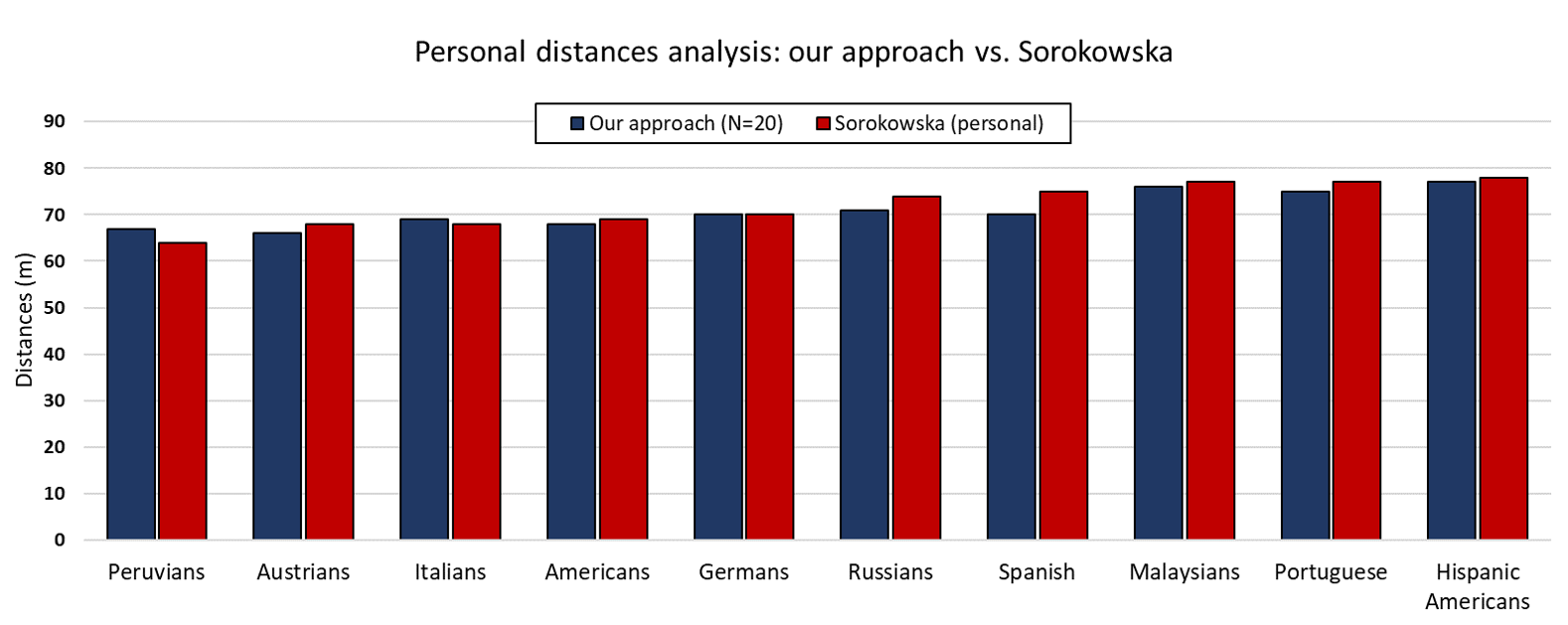}
  \caption{Comparison between personal distances obtained in our method and Sorokowska et al.~\cite{sorokowska:2017} work. Such data was obtained simulating $20$ agents in the virtual FD and compared with Sorokowska et al.~\cite{sorokowska:2017} results in 10 Countries.}
    \label{fig:fd:soroka_compare}
\end{figure}

Along with the comparison between real crowds and literature for Countries Brazil and Germany, we executed simulations for other cultural groups (Countries) represented both into Sorokowska et al.~\cite{sorokowska:2017} and McCrae~\cite{OCEAN_GLOBAL_COMPARISON} studies. By comparing the simulated personal distances, obtained using McCrae~\cite{OCEAN_GLOBAL_COMPARISON} cultural OCEAN as input, with Sorokowska et al.~\cite{sorokowska:2017} results, we formatted the obtained results in Figure~\ref{fig:fd:soroka_compare}.

As showed in Figure~\ref{fig_our_vs_soro}, we compare our results for experiments with $15$, $20$, and $25$ pedestrians with Sorokowska's et al. results. It is interesting to see that the values are much more similar with $30$ pedestrian than with~$15$. This is explained because the results obtained in Sorokowska et al.~\cite{sorokowska:2017} work are related to personal distances, according to Hall~\cite{HALL1957theories}, i.e. values from~$45cm$ to~$120cm$. So, in a simulated in the environment with~$30$ agents, we present the situation where people are in personal distances and results can be compared with Sorokowska's et al.

Regarding the Figure~\ref{fig:fd:soroka_compare}, it is easy to see that in some Countries population, e.g. from Peruvians to Hispanic Americans (in X axes), the values of personal spaces are very similar.
It indicates that the input according to McCrae~\cite{OCEAN_GLOBAL_COMPARISON} OCEAN values is correlated with the physical space occupied by agents in our simulation, when compared to real pedestrians.

\section{Discussions and Final Considerations}
\label{sec:conclusion}

In this paper we presented some comparatives in cultural aspects of group of people in video sequences from two countries: Brazil and Germany. Since one important aspect to be considered in behavior analysis is the context and environment where people are acting, we worked with Fundamental Diagram experiment proposed by~\cite{Chattaraj:2009}, in this way, people from both countries performed exactly the same task. Our hypothesis is that by fixing the environment setup and the task people should apply, we could evaluate the cultural variation of individual behavior.

In the analysis, we found out that as the density of people increases, people are more homogeneous, as shown in PDF of distances (Figure~\ref{fig_pdf}) and Kullback-Leibler divergence in Figure~\ref{fig_kld} and in computed Pearson's correlation in Figure~\ref{fig_corr_dis}. It indicates that people assumes group-level behavior instead of individual-level behavior according to his/her culture or personality. It is an interesting and concrete proof of several theories about mass behavior as discussed in Vilanova~\cite{vilanova2017} and Le Bon~\cite{LE_BON_THE_CROWD}.

We show some differences among Brazil and Germany in the personal space of the individuals in terms of distances between individuals. These differences are evidences of cultural behavior of people from each country, mainly in low density or small groups, when the individuals are not acting as a crowd.

We performed a comparison among the personal spaces pedestrian keep from others in the videos of FD with the study proposed by Sorokowska~\cite{sorokowska:2017}. It was interesting to see that the personal spaces observed in the videos from Brazil and Germany in FD experiment are are in accordance with those presented through subject answers given in the Sorokowska's work.

In addition, we proposed a way to simulate the FD experiment from other countries. For this, we use the OCEAN of each country as input to discover the collectivity, angular variation and linear speed of each agent in the simulation. We also used Sorokowska distances to compare the distances between agents obtained in the simulations of each country. The results also are in accordance whit Sorokowska~\cite{sorokowska:2017}.

For future work, we intend to keep investigating the cultural aspects in video sequences, focused on medium and low densities, since it seems to be more different in terms of culture at this densities of pedestrians. We also intend to increase our set of video data, addressing another countries.


\section*{References}

\bibliography{references}

\end{document}